\documentclass[reprint,amsmath,amssymb,aps,biblatex,prx,superscriptaddress]{revtex4-2}

\usepackage{graphicx}
\usepackage{dcolumn}
\usepackage{bm}
\usepackage{lineno}
\usepackage{braket}
\usepackage{svg}
\usepackage{siunitx}
\usepackage{bbm}
\usepackage{multirow}
\usepackage{booktabs}
\usepackage{xcolor}

\newcommand*\diff{\mathop{}\!\mathrm{d}}

\begin{document}

\title{Analog quantum simulation of the Lipkin-Meshkov-Glick model in a transmon qudit}

\author{Elizabeth Champion}
\author{Annie Schwartz}
\author{Muhammad A. Ijaz}
\author{Xiaohui Xu}
    \affiliation{Department of Physics and Astronomy, University of Rochester, Rochester, New York 14627, USA}
    \affiliation{University of Rochester Center for Coherence and Quantum Science, Rochester, New York 14627, USA}
\author{Steve Campbell}
    \affiliation{School of Physics, University College Dublin, Belfield, Dublin 4, Ireland}
    \affiliation{Centre for Quantum Engineering, Science, and Technology, University College Dublin, Dublin 4, Ireland}
\author{Gabriel T. Landi}
\author{Machiel S. Blok}
    \email{machielblok@rochester.edu}
    \affiliation{Department of Physics and Astronomy, University of Rochester, Rochester, New York 14627, USA}
    \affiliation{University of Rochester Center for Coherence and Quantum Science, Rochester, New York 14627, USA}

\date{\today}

\begin{abstract}
\noindent The simulation of large-scale quantum systems is one of the most sought-after applications of quantum computers.
Of particular interest for near-term demonstrations of quantum computational advantage are analog quantum simulations, which employ analog controls instead of digitized gates. 
Most analog quantum simulations to date, however, have been performed using qubit-based processors, despite the fact that many physical systems are more naturally represented in terms of qudits (i.e., $d$-level systems).
Motivated by this, we present an experimental realization of the  Lipkin-Meshkov-Glick (LMG) model using an analog simulator based on a single superconducting transmon qudit with up to $d = 9$ levels.
This is accomplished by moving to a rotated frame in which evolution under any time-dependent local field and one-axis twisting can be realized by the application of multiple simultaneous drives.
Combining this analog drive scheme with universal control and single-shot readout of the qudit state, we provide a detailed study of five finite-size precursors of quantum criticality in the LMG model: dynamical phase transitions, closing of the energy gap, Kibble-Zurek-like dynamics, statistics of the order parameter, and excited-state phase transitions.
For each experiment we devise a protocol for extracting the relevant properties which does not require any prior knowledge of the system eigenstates, and can therefore be readily extended to higher dimensions or more complicated models.
Our results cement high-dimensional transmon qudits as an exciting path towards simulating many-body physics.
\end{abstract}

\maketitle

Simulating the dynamics of large quantum systems is a promising application for quantum computation \cite{georgescu_quantum_2014, daley_practical_2022}.
Recent qubit-based experiments explored a range of interesting dynamical quantum phenomena including topological order in strongly interacting fermions \cite{evered_probing_2025}, string breaking in lattice gauge theories \cite{gonzalez-cuadra_observation_2025}, and quantum criticality in spin lattices \cite{andersen_thermalization_2025}.
Ultimately, the goal of quantum simulation is to emulate classically intractable problems with physical relevance.
However, not all physically relevant systems are easily emulated using qubits.
Many physical systems of interest have higher-dimensional local Hilbert spaces, and encoding a high-dimensional quantum system into multiple qubits requires engineering complex (non-local) multi-body interactions, introducing significant overhead.
These challenges suggest that exciting new opportunities may emerge from exploring quantum evolution in qudit-based simulators \cite{ciavarella_trailhead_2021, gustafson_prospects_2021, gonzalez-cuadra_hardware_2022, zache_fermion-qudit_2023, illa_qu8its_2024, meth_simulating_2025, calliari_quantum_2025,macdonell_analog_2021,pausch_dissipative_2024}, in which two-level qubits are replaced by $d$-level quantum systems.

An ideal subject of study for qudit-based simulators is the Lipkin-Meshkov-Glick (LMG) model \cite{lipkin_validity_1965}. 
This is a widely used model in many-body physics, whose large-spin Hamiltonian is naturally encoded into a qudit.
The LMG model hosts a rich variety of quantum critical phenomena \cite{castanos_classical_2006, ribeiro_thermodynamical_2007, ribeiro_exact_2008} owing to a competition between the local field ($J_z$) and one-axis-twisting (OAT, $J_x^2$) terms.
Despite rapid progress in the development of qudit architectures for quantum information processing \cite{ringbauer_universal_2022, low_control_2025,yu_schrodinger_2025, vaartjes_certifying_2025,blok_quantum_2021, morvan_qutrit_2021, goss_high-fidelity_2022, nguyen_empowering_2024, liu_performing_2023, wang_high-e_je_c_2025} and quantum simulation \cite{senko_realization_2015, edmunds_symmetry-protected_2025, meth_simulating_2025,wang_dissipative_2023, kumaran_transmon_2025, wang_observing_2025}, a qudit simulator with tunable local fields and OAT has thus far not been realized.
A static qudit-OAT term can arise naturally in nuclear spins through the quadrupole splitting \cite{yu_schrodinger_2025, vaartjes_certifying_2025} or can be mapped to a Kerr nonlinearity in transmons \cite{champion_efficient_2025}.
For these systems the OAT term provides the nonlinearity that enables universal control, but it is challenging to tune the term in-situ because resonant control fields do not obey the parity symmetry of the OAT term.
A realization of programmable OAT in qudits would be a powerful addition to the simulator toolbox for studying  dynamical qudit phenomena \cite{kitagawa_squeezed_1993}.

In this work we engineer local fields and one-axis twisting in a high-$E_J / E_C$ superconducting transmon qudit with up to $d = 9$ levels \cite{wang_high-e_je_c_2025} to study quantum criticality in the LMG model.
Analog evolution under the LMG Hamiltonian is realized in a rotating frame by applying simultaneous drive tones close to each qudit transition \cite{champion_efficient_2025}, where the detunings and amplitudes of the drives are independently varied to parametrize the local field and OAT.
We furthermore develop an adiabatic protocol that facilitates the preparation of LMG eigenstates and their superpositions without requiring any prior knowledge of the Hamiltonian's eigenstates -- a crucial condition for future quantum simulators.
With these tools, we experimentally investigate a total of five aspects of the LMG criticality: (1) dynamical phase transitions, (2) closing of the energy gap, (3) Kibble-Zurek-like dynamics, (4) statistics of the order parameter, and (5) excited-state phase transitions.
Our experiments involve spin sizes below the scaling region, and therefore shed light on the finite-size precursors of these phase transitions, and how they can be probed experimentally in a system with universal quantum control.
In conjunction with recent work on transmon qudit entangling gates \cite{morvan_qutrit_2021, goss_high-fidelity_2022, nguyen_empowering_2024}, our results lay the groundwork for quantum simulation of complex qudit-based Hamiltonians with future applications in particle physics \cite{ciavarella_trailhead_2021, gustafson_prospects_2021, gonzalez-cuadra_hardware_2022, zache_fermion-qudit_2023, illa_qu8its_2024, meth_simulating_2025, calliari_quantum_2025}, quantum chemistry \cite{macdonell_analog_2021}, and condensed matter \cite{pausch_dissipative_2024}.\\

\begin{figure*}
    \includegraphics{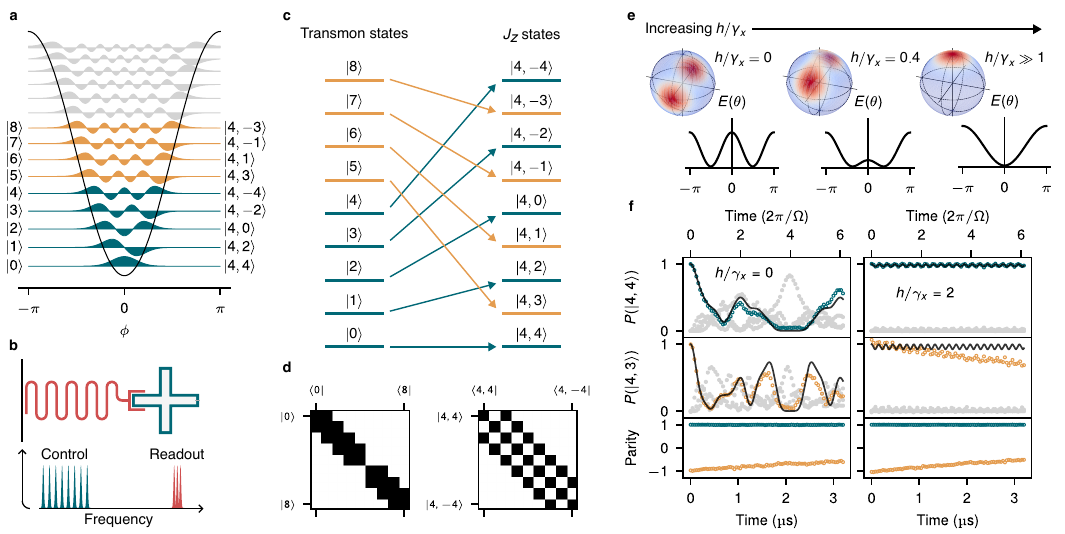}
    \caption{Analog quantum simulation of the LMG model.
    \textbf{a}, High-$E_J / E_C$ transmon potential well and eigenstate wavefunctions. The state labels on the left indicate the physical transmon states, while those on the right indicate the encoded spin states.
    \textbf{b}, Diagram of the transmon and resonator (top) and the drive spectrum used for control and readout (bottom, not to scale). Multiple simultaneous drive tones are used to simulate the LMG Hamiltonian in a rotating frame, and three simultaneous readout tones enable single-shot measurement of the qudit state.
    \textbf{c}, Diagram of the state labeling scheme. The even-parity spin states are encoded into the lowest $j + 1$ levels of the transmon, while the odd-parity spin states are encoded into the remaining $j$ levels.
    \textbf{d}, Nonzero matrix elements of the LMG Hamiltonian in the transmon basis (left) and the $J_z$ basis (right). In the $J_z$ basis the Hamiltonian has nonzero elements on the diagonal and second off-diagonals, while in the transmon basis, it has elements on the diagonal and first off-diagonals.
    \textbf{e}, Symmetry-broken and normal phases of the LMG model. We show the Husimi $Q$ functions of the LMG ground states for $j = 4$ with $h / \gamma_x = 0$ (top left), $h / \gamma_x = 0.4$ (top middle), and $h / \gamma_x \gg 1$ (top right). The corresponding semiclassical energy surfaces are shown in the bottom row.
    \textbf{f}, Experimentally measured evolution of the $J_z$ eigenstates $\ket{4, 4}$ (top) and $\ket{4, 3}$ (middle) under the $j = 4$ LMG Hamiltonian with $h / \gamma_x = 0$ (left) and $h / \gamma_x = 2$ (right). Colored points give the measured populations in the initial states, solid lines indicate the theoretical evolution, and gray points give the other measured populations within the same-parity subspaces. The bottom panels show the expectation value of the parity operator in each case, showing that parity is preserved up to $T_1$ decay from the odd subspace into the even subspace.
    }
    \label{fig:analog_sim}
\end{figure*}

The LMG model was first introduced as a testbed for many-body approximation methods in the context of nuclear physics because it has exact analytical solutions in some cases, yet features interesting quantum phenomena \cite{lipkin_validity_1965}.
The model can be formulated as $N$ degenerate, all-to-all interacting spin-$1/2$ particles subject to an external field, with Hilbert space dimension $2^N$.
However, the structure of the model permits a collective spin representation in which the highest-spin sector has size $j = N / 2$ (for simplicity, in this work we consider integer $j$).
In this picture the Hamiltonian is given by
\begin{equation}
    H / \Omega = -h J_z - \frac {\gamma_x} {2j} J_x^2 - \frac {\gamma_y} {2j} J_y^2,
\end{equation}
where $h$ is the field strength, $\gamma_x$ and $\gamma_y$ describe the twisting terms (corresponding to interaction terms in the many-body model), and $\hbar= 1$.
We take the parameters $h$, $\gamma_x$, and $\gamma_y$ to be dimensionless, and fix the simulation energy scale to $\Omega / (2 \pi) = \qty{1.910}{\mega\hertz}$ unless stated otherwise.
Here we focus on the one-axis twisting case, where $\gamma_y = 0$,
\begin{equation} \label{eq:LMG}
    H_{\rm LMG} / \Omega = -h J_z - \frac {\gamma_x} {2j} J_x^2,
\end{equation}
although we note that the inclusion of the $J_y^2$ term (leading to two-axis twisting) is straightforward in our experimental setup.
We denote $J_z$ eigenstates by $\ket{j,m}$ for $-j \leq m \leq j$ such that $J_z \ket{j,m} = m \ket{j,m}$.
The eigenstates of $J_x$ are denoted by $\ket{j,m}_x$ such that $J_x \ket{j,m}_x = m \ket{j,m}_x$.

The Hamiltonian~(\ref{eq:LMG}) is invariant under the $\mathbb{Z}_2$ transformation $(J_x,J_y,J_z) \rightarrow (-J_x,-J_y,J_z)$, i.e., a rotation by $\pi$ about the $z$ axis.
We define a parity operator $\Pi = \exp \left [ i \pi (J_z - j) \right ]$, which has degenerate eigenvalues $\pm 1$ corresponding to even and odd parities respectively.
The additional factor of $e^{-i \pi j}$ ensures that the ground state is always in the even-parity sector.
Because $H_{\rm LMG}$ commutes with $\Pi$, they share an eigenbasis.
In the limit where $h/\gamma_x \gg 1$ the definite-parity eigenstates are the $J_z$ eigenstates, while for $h/\gamma_x \ll 1$ they are given by $\ket{j, m}_\pm = \frac {1} {\sqrt{2}} (\ket{j,m}_x \pm \ket{j,-m}_x)$ when $m \neq 0$, and $\ket{j, 0}_+ = \ket{j, 0}_x$; see the Supplemental Material for details.

We can develop intuition for the LMG model through a semiclassical analysis~\cite{ribeiro_exact_2008, castanos_classical_2006} where $J_{x,y,z}$ are imagined as the components of a vector on a sphere of radius $j$,
\begin{equation}\label{eq:J}
    \mathbf{J} = j (\sin\theta \cos\phi, \sin\theta \sin\phi, \cos\theta),
\end{equation}
with classical energy surface
\begin{equation}
    \frac {E(\theta, \phi)} {j} = -h \cos\theta - \frac {\gamma_x} {2} \sin^2\theta \cos^2\phi.
\end{equation}
For $h/\gamma_x \geq 1$ this has a single minimum at $\theta_0 = 0$, and this region is henceforth referred to as the normal phase.
For $h / \gamma_x < 1$ OAT dominates, leading to two degenerate minima located at $\theta_0 = \pm\arccos(h/\gamma_x)$ and $\phi = 0$, and we call this region the symmetry-broken phase (see the Supplemental Material for details).
The energy surfaces are shown in Figure \ref{fig:analog_sim}e, along with Husimi $Q$ functions for the $j = 4$ ground states.

In this work we experimentally investigate the properties of the LMG model using a fixed-frequency transmon qudit of dimension $d = 2j + 1$.
All experiments were carried out using the transmon labeled $Q_5$ in Ref. \cite{wang_high-e_je_c_2025}, which was designed to have $E_J / E_C \approx 300$ \cite{koch_charge-insensitive_2007, blais_circuit_2021}.
This ensures that a large number of states are confined in the cosine potential, and strongly suppresses charge dispersion for all states used in this work (Figure \ref{fig:analog_sim}a).
By selectively driving individual transitions, we can prepare arbitrary qudit states and perform arbitrary unitary rotations with high fidelity \cite{morvan_qutrit_2021, liu_performing_2023, wang_high-e_je_c_2025, champion_efficient_2025}.
Meanwhile, a dispersively coupled readout resonator facilitates single-shot readout of the full qudit state (Figure \ref{fig:analog_sim}b).
We provide device parameters and coherence time measurements in the Supplemental Material, and refer the reader to Ref. \cite{wang_high-e_je_c_2025} for a detailed discussion of the device properties and the multi-tone readout protocol.

The LMG Hamiltonian is simulated by applying simultaneous drives with frequencies near the transmon's single-photon transitions.
Denoting the transmon eigenstates by $\ket{n}$ for $0 \leq n < d$, we show (see Methods) that moving to a rotating frame that is detuned from the bare transmon frame and taking the rotating wave approximation (RWA) yields a Hamiltonian of the form 
\begin{equation} \label{eq:rwa_hamiltonian}
    H(t) = \sum_{n=0}^{d-1} \Delta_n(t) \ket{n} \bra{n} + \sum_{n=1}^{d-1} \bigg ( \Omega_n(t) \ket{n-1} \bra{n} + \text{h.c.} \bigg ),
\end{equation}
where $\Omega_n(t)$, $\Delta_n(t)$ are the amplitudes and detunings of the drives that appear on the first off-diagonal and diagonal, respectively (Figure \ref{fig:analog_sim}d, left).
Meanwhile, writing the LMG Hamiltonian in the $J_z$ basis, we find that it contains nonzero matrix elements only on the diagonal and second off-diagonals, as depicted in the right panel of Figure \ref{fig:analog_sim}d. 
Importantly, however, we are free to encode the Hilbert space of the emulated spin into the transmon Hilbert space in whichever manner is most convenient.
As shown in Figure \ref{fig:analog_sim}c, we encode the even-parity $J_z$ eigenstates in the lowest $j + 1$ transmon eigenstates, and the odd-parity $J_z$ eigenstates in the remaining $j$ transmon states.
This relabeling of the Hilbert space has the effect of moving the second-off-diagonal (OAT) elements onto the first off-diagonals (Figure \ref{fig:analog_sim}d, left), thereby allowing for the direct analog simulation of the LMG model.
The drive amplitudes, detunings, and phases are determined by computing $H_{\rm LMG}$ in this relabeled basis and comparing to Eq. \ref{eq:rwa_hamiltonian}.

We now present the results of several experiments which probe the finite-size precursors of quantum criticality in the LMG model.
We give a detailed explanation of the protocol for each experiment in the Methods section.

\begin{figure*}
    \includegraphics{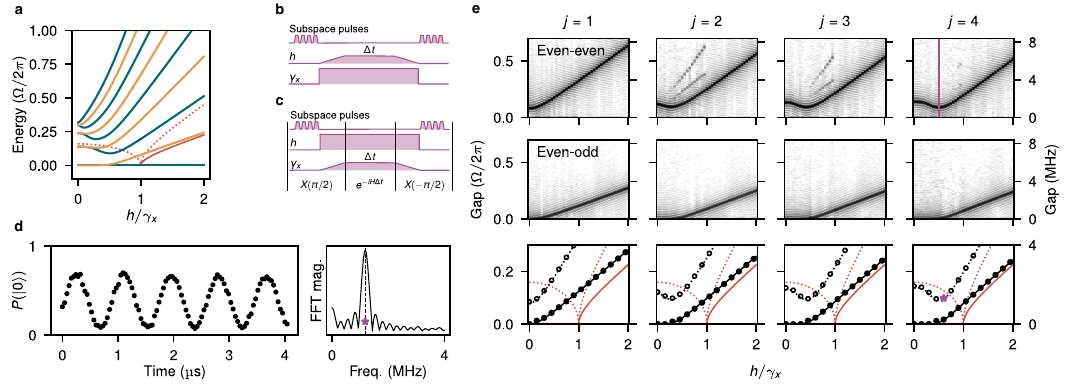}
    \caption{Adiabatic protocol and gap measurement.
    \textbf{a}, Energy spectrum of the LMG model with $j = 4$ as a function of $h / \gamma_x$. The even-parity states are shown in blue, and odd-parity in yellow. In this experiment we focus on the lowest even-odd and even-even energy gaps. These gaps in the thermodynamic limit are shown in red.
    \textbf{b}, Gap measurement protocol for $h / \gamma_x < 1$.
    \textbf{c}, Gap measurement protocol for $h / \gamma_x > 1$. Both protocols implement Ramsey-like experiments in which we adiabatically prepare equal superpositions of LMG eigenstates, evolve under the LMG Hamiltonian, and project the resulting state onto the initial state.
    \textbf{d}, Time trace (left) and Fourier transform (right) for a single gap experiment with $j = 4$ and $h / \gamma_x = 0.6$. This corresponds to the purple line in the top right panel of \textbf{e}, and the peak of the Fourier transform gives the energy indicated by the purple star in the bottom right panel of \textbf{e}.
    \textbf{e}, Experimental results for the even-even (top) and even-odd (middle) gaps across a range of $h / \gamma_x$ values. The plots are obtained by computing the Fourier transform of each Ramsey signal as shown in \textbf{d} and plotting the logarithm of its magnitude in grayscale. We see strong peaks corresponding to the gap. In the bottom row we extract the frequencies of these peaks (open circles for even-even, filled circles for even-odd) and plot them on top of the theoretically expected values (dashed black line for even-even, solid black line for even-odd). We also show the corresponding gaps in the thermodynamic limit, $j \rightarrow \infty$, in red.
    }
    \label{fig:gap}
\end{figure*}

\subsection*{Dynamical phase transition}

We demonstrate the time evolution of the system under the LMG Hamiltonian for $j = 4$ in Figure \ref{fig:analog_sim}f, where in the top row we begin in the state $\ket{4,4}$ and in the middle row we begin in $\ket{4,3}$.
The left column corresponds to evolution in the symmetry-broken phase ($h = 0$) and the right column in the normal phase ($h = 2$); in both cases we set $\gamma_x = 1$.
The colored points represent the Loschmidt Echo (return probability) and the gray dots are the occupation probabilities for the other $J_z$ eigenstates of the same parity.
When $h = 0$ the initial states are not close to eigenstates of the Hamiltonian, so we observe that the system quickly evolves away from its initial state, in agreement with the theoretical predictions (solid lines).
The system undergoes coherent oscillations between $J_z$ eigenstates (Figure \ref{fig:analog_sim}f, left) and rapidly evolves to a state nearly orthogonal to its initial state, representing a finite-size precursor of the dynamical phase transition~\cite{heyl_dynamical_2013, heyl_dynamical_2014,jafari_loschmidt_2017,cejnar_impact_2008, puebla_finite-component_2020, santos_structure_2015} and the so-called orthogonality catastrophe~\cite{campbell_criticality_2016,fogarty_orthogonality_2020}.
This behavior is also closely related to the LMG model's excited-state quantum phase transition, discussed further below.
Meanwhile, when $h = 2$ (Figure \ref{fig:analog_sim}f, right column) the LMG eigenstates are close to those of $J_z$, causing the return probabilities to remain close to $1$.
In the bottom row of Figure \ref{fig:analog_sim}f we plot the expectation value of the parity operator for each case, computed from the measured populations.
The parity should ideally be $\langle \Pi \rangle = \pm 1$ for all times, because we begin in states of definite parity.
For the even-parity case the parity remains essentially unchanged during the dynamics, while for odd-parity case (beginning in $\ket{4,3}$) we observe that the parity slowly evolves away from $-1$, owing to gradual $T_1$ decay of the transmon from the odd-parity subspace into the even-parity subspace.
This experiment thus showcases the ability of our analog simulator to preserve the underlying symmetries of the model.

\subsection*{Closing of the ground-state gap}

In our next experiment we measure the LMG model's energy spectrum, focusing on gaps between low-energy states within the same-parity subspace as well as between subspaces.
We show the spectrum of the system for $j = 4$ as a function of $h / \gamma_x$ in Figure \ref{fig:gap}a.
We note that, in the thermodynamic limit, the closing of the energy gaps at $h / \gamma_x = 1$ (see, e.g., the red curves in Figure \ref{fig:gap}a) is a signature of the transition between the symmetric and symmetry-broken phases.
Here we measure the energy gaps between the ground state (even-parity) and the first and second excited states, which have odd and even parities, respectively.
We denote these, respectively, by $\ket{0_{\rm LMG}}$, $\ket{1_{\rm LMG}}$, and $\ket{2_{\rm LMG}}$.
We measure the energy gaps $E_{\rm even-odd} = E_{\ket{1_{\rm LMG}}} - E_{\ket{0_{\rm LMG}}}$ and $E_{\rm even-even} = E_{\ket{2_{\rm LMG}}} - E_{\ket{0_{\rm LMG}}}$ by performing Ramsey-style experiments in which we prepare equal superpositions of the states of interest and evolve under the LMG Hamiltonian for a variable time $\Delta t$, imparting a phase difference $-E \Delta t$ between the eigenstates.
Projecting the final state onto the initial superposition results in an oscillation at a frequency proportional to the energy gap.
To avoid requiring \emph{a priori} knowledge of the LMG eigenstates (which would be an issue in more complicated models), we modify this protocol to include adiabatic state preparation.
We start by preparing equal superpositions of LMG eigenstates in the limits $h/\gamma_x \ll 1$ and $h / \gamma_x \gg 1$, for which the eigenstates are trivially known, and subsequently ramp the LMG Hamiltonian parameters to their desired values, as depicted in Figures \ref{fig:gap}b and \ref{fig:gap}c.
The initial states and the direction of the ramp are chosen so as to avoid ramping the system through the critical point at $h / \gamma_x = 1$.
The result is an equal superposition of LMG eigenstates for arbitrary $h / \gamma_x$, provided that the ramp is slow enough to satisfy the adiabatic approximation (see Methods).

We determine the gap energy from the Ramsey oscillations by computing the Fourier transform of the time-domain signal measured in the experiment, as shown in the right panel of Figure \ref{fig:gap}d.
We see a strong peak in the spectrum, which agrees well with the theoretical value for the gap (vertical dashed line).
In Figure \ref{fig:gap}e we show the results for the even-odd and even-even gaps for $j = 1$ through $j = 4$ across the range $0 \leq h/\gamma_x \leq 2$, where the purple line in the top right panel corresponds to the signal plotted in Figure \ref{fig:gap}d.
In each case we see a strong peak corresponding to the gap energy.
These peaks are extracted and plotted on top of the theoretical gap curves in the bottom panel of Figure \ref{fig:gap}e, and in all cases we see very good agreement between theory and experiment without any fit parameters.
As the size of the system is increased, the gaps begin to more closely resemble the theoretical gaps in the thermodynamic limit (red curves in the bottom row of Figure \ref{fig:gap}e).
In some cases (for example, the even-even gaps for $j = 2$ and $j = 3$ in Figure \ref{fig:gap}e), we also see secondary peaks in the Fourier transform.
Because we are simulating a low-dimensional realization of the LMG model, the phase transition at $h/\gamma_x = 1$ is replaced by a minimum in the even-even gap as well as a splitting of the approximate degeneracy of the even-odd gap in the region $h/\gamma_x < 1$ (Figure \ref{fig:gap}a).
When ramping $h/\gamma_x$ from 0 to a nonzero value above this minimum-gap point, we see spurious excitations arising from a breakdown of the adiabatic approximation, whose exact amplitudes depend upon the detailed dynamics of the system as it passes through this point.

\subsection*{Kibble-Zurek-like dynamics}

\begin{figure}
    \includegraphics[width=\linewidth]{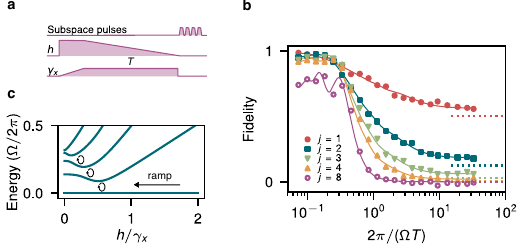}
    \caption{Transition between adiabatic and impulse regimes.
    \textbf{a}, Experimental protocol for the measurement. We adiabatically prepare the $h / \gamma_x = 2$ ground state, then ramp $h$ to zero over a variable time $T$.
    \textbf{b}, Experimental results showing the probability that the system ends in the $h / \gamma_x = 0$ eigenstate as a function of ramp speed. The solid curves are single-parameter fits of numerical simulations to the data, where the the effect of decoherence is modeled as an exponential decay of the measured population. The dotted lines show the overlaps between the initial and final LMG eigenstates.
    \textbf{c}, Energy spectrum of the even-parity LMG states. When the system is ramped non-adiabatically through the minimum-gap point, it undergoes transitions to the subspace of even-parity excited states.
    }
    \label{fig:ground_state}
\end{figure}

We study the creation of excitations (defects) as the system is driven across the critical point by a non-adiabatic drive.
Building on insights related to the formation of topological defects, Zurek put forward predictions for how the number of excitations should scale with the ramp speed~\cite{zurek_cosmological_1985, zurek_dynamics_2005,dziarmaga_dynamics_2005, damski_simplest_2005}. 
The LMG model does not generally follow this scaling for arbitrary ramps, owing to its long-range (all-to-all) interaction~\cite{defenu_dynamical_2018}. 
Notwithstanding, Kibble-Zurek-like experiments remain a valuable tool to probe the non-equilibrium response to fast drives.
We investigate this excitation process in Figure \ref{fig:ground_state}b for $j = 1$ through $j = 4$ as well as $j = 8$ (discussed further below).
The protocol, shown in Figure \ref{fig:ground_state}a and detailed in the Methods section, begins by adiabatically preparing the LMG ground state for $h / \gamma_x = 2$.
The system is then ramped to $h / \gamma_x = 0$ over a variable time $T$, after which we measure the population remaining in the instantaneous ground state.
We plot this measured population as a function of the ramp speed, $2\pi / (\Omega T)$.
For slow ramps we see that the system ends in the final ground state with high fidelity, with an overall loss of contrast for larger spins due to decay.
For fast ramps, meanwhile, the overlap with the final ground state is low.
As illustrated in Figure \ref{fig:ground_state}c, as the system is ramped through the point where the gap reaches a minimum, it can be sequentially excited from the ground state to the subspace of even-parity excited states.
For very fast ramps, the fidelity saturates to the overlap between the initial and final LMG eigenstates, indicated by the dotted lines on the right side of Figure \ref{fig:ground_state}b.
This behavior illustrates how rapid quenches can cause diabatic transitions out of the instantaneous ground state.

\begin{figure*}
    \includegraphics{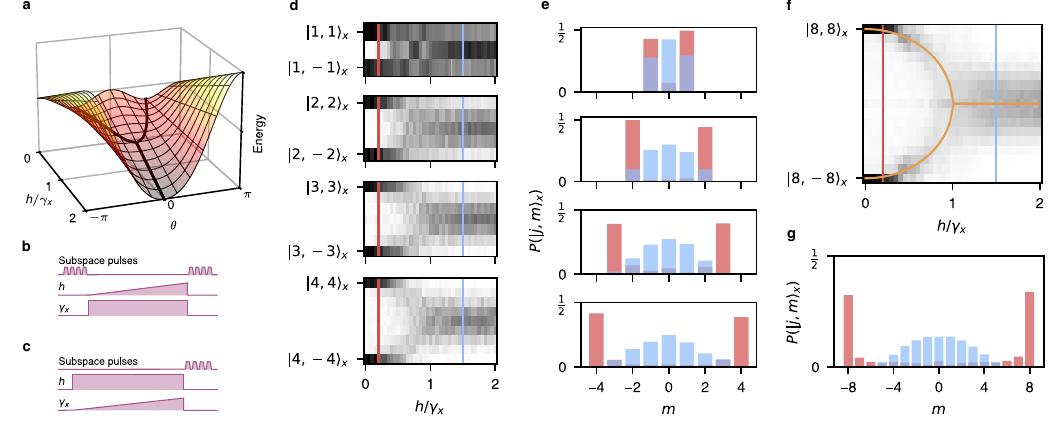}
    \caption{Distribution of $J_x$ eigenstates in the LMG ground state.
    \textbf{a}, Semiclassical energy surface as a function of $\theta$ and $h / \gamma_x$, setting $\phi = 0$. In the symmetry-broken phase ($h / \gamma_x < 1$), the energy surface has two degenerate minima. At the critical point ($h / \gamma_x = 1$) these minima merge, and we have a single minimum for $h / \gamma_x > 1$.
    \textbf{b}, Experimental protocol for $h / \gamma_x < 1$.
    \textbf{c}, Experimental protocol for $h / \gamma_x > 1$. In both cases we adiabatically prepare the LMG ground state as described above, then perform a series of pulses mapping one of the $J_x$ eigenstates onto the transmon ground state prior to readout.
    \textbf{d}, Histograms representing the $J_x$ eigenstate distributions as a function of $h / \gamma_x$ for $j = 1$ through $j = 4$.
    \textbf{e}, $J_x$ eigenstate distributions for the particular values of $h / \gamma_x$ indicated by the vertical lines in \textbf{d}.
    \textbf{f}, Histograms representing the $J_x$ eigenstate distributions as a function of $h / \gamma_x$ for $j = 8$, measured by projecting the $J_x$ eigenstates onto the even-parity subspace and renormalizing. As discussed in the text, the yellow curves give the locations of the minima of the classical energy surface when projected onto the $J_x$ axis.
    \textbf{g}, $J_x$ eigenstate distributions for the particular values of $h / \gamma_x$ indicated by the vertical lines in \textbf{f}.
    }
    \label{fig:order}
\end{figure*}

The data shown in Figure \ref{fig:ground_state}b for $j = 8$ are measured by observing that, if we are only concerned with the dynamics within a subspace of definite parity, there is no need to simulate the opposite-parity subspace.
The LMG ground state always has even parity, and because the LMG Hamiltonian is parity-preserving, the state must remain within the even-parity subspace for the duration of the experiment, even if it is excited out of the instantaneous LMG ground state.
Therefore, we can double the size of the effective spin in any experiment where we are solely concerned with the even- or odd-parity subspace.
We make further use of this idea below.

\subsection*{Statistics of the order parameter}

We next turn to the structure of the LMG ground state.
The order parameter is $\langle J_x \rangle$.
However, for any finite size this will always vanish due to symmetry.
To visualize the emergence of the symmetry-broken phase as the spin size increases, we therefore measure the full probability distribution $|\braket{0_{\rm LMG} | j,m}_x |^2$ of $J_x$, in the LMG ground state. 
We can borrow intuition from the classical energy surface of Figure \ref{fig:order}a.
In the normal phase ($h / \gamma_x > 1$) there is a single minimum at $\theta_0 = 0$, while in the symmetry-broken phase ($h / \gamma_x < 1$) there are two degenerate minima located at $\theta_0 = \pm\arccos(h/\gamma_x)$.
Since classically [Eq.~\eqref{eq:J}] $J_x \sim j \sin\theta$, we expect that $|\braket{0_{\rm LMG} | j,m}_x |^2$ should be peaked around $m(\theta_0) = j \sin\theta_0$, and the transition from a unimodal to a bimodal distribution can be taken as a finite-size signature of the broken symmetry. 
By measuring the distribution of $J_x$ eigenstates we can therefore directly compare the quantum and semiclassical pictures.

The protocol for this measurement in the case where $h/\gamma_x < 1$ is shown in Figure \ref{fig:order}b, while that for $h/\gamma_x > 1$ is shown in Figure \ref{fig:order}c.
In either case, we adiabatically prepare the LMG ground state in a manner analogous to that discussed above, then apply a series of pulses mapping a single $J_x$ eigenstate $\ket{j,m}_x$ onto the transmon ground state.
Measuring the population in the transmon ground state yields the overlap $|\braket{0_{\rm LMG} | j,m}_x |^2$, and we repeat this for all $\ket{j,m}_x$ to measure the full distribution.
In Figure \ref{fig:order}d we show the distribution across a range of $h/\gamma_x$, while in Figure \ref{fig:order}e we show it for two specific $h / \gamma_x$ values, indicated by the vertical lines in Figure \ref{fig:order}d.
As $h/\gamma_x \rightarrow 0$, recall that the ground state becomes $\ket{j, j}_+ = \frac {1} {\sqrt{2}} (\ket{j,j}_x + \ket{j,-j}_x)$.
We therefore see two peaks in the eigenstate distribution at its extreme values.
As $h/\gamma_x$ is increased these peaks begin to merge, converging to a single peak centered around $\ket{j,0}_x$.
This can be understood as the ground state approaching the $-J_z$ ground state, $\ket{j,j}$: in the $J_x$ basis, this state is a spin coherent state on the equator of the rotated phase space sphere.

In Figures \ref{fig:order}f and \ref{fig:order}g we measure the $J_x$ eigenstate distribution for $j = 8$ (see Methods for details).
In Figure \ref{fig:order}f we also plot the locations of the minima of the classical energy surface projected onto the $J_x$ axis, $m(\theta_0)$, as discussed above.
The semiclassical intuition is reproduced in this case, in the sense that the centers of the peaks in the distribution qualitatively agree with the semiclassical prediction.
If we were to increase $j$ further, the relative widths of the peaks would decrease, ultimately approaching delta functions centered on the locations predicted by the semiclassical analysis.

\begin{figure}
    \includegraphics{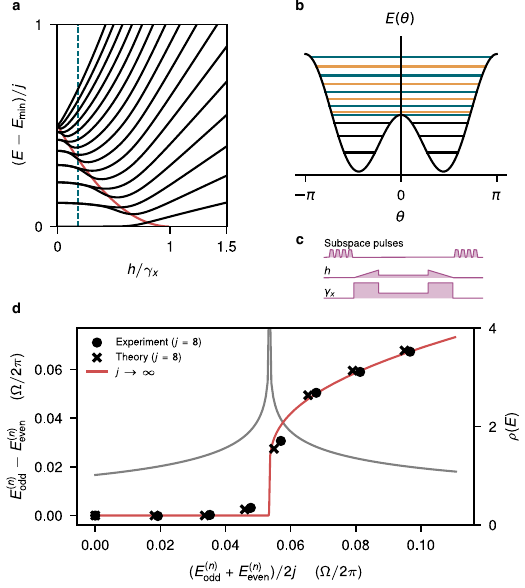}
    \caption{Excited-state quantum phase transition (ESQPT).
    \textbf{a}, LMG energy spectrum as a function of $h / \gamma_x$ for $j = 8$. The red curve indicates the critical energy $E_c(h)$ of the ESQPT, equal to the height of the barrier of the classical energy surface relative to its minimum energy.
    \textbf{b}, Semiclassical energy surface for $h / \gamma_x = 0.18$ (vertical dashed line in \textbf{a}), with $\phi = 0$. The horizontal lines show the $j = 8$ eigenenergies. For energies near or above $E_c(h)$, we plot even-parity eigenenergies in blue and odd-parity eigenenergies in yellow.
    \textbf{c}, Experimental protocol. We measure energy gaps between adjacent states of like parity in a manner analogous to the gap measurement in Figure \ref{fig:gap}. Here, however, we use a different energy scale $\Omega$ during the ramps in order to more easily satisfy the adiabatic approximation (see Methods).
    \textbf{d}, Experimental results for the energy difference of each even-odd pair as a function of that pair's average energy, along with the theoretical expectation for $j = 8$, as measured here, and $j \rightarrow \infty$. We also show the semiclassical density of states result in gray.
    }
    \label{fig:excited}
\end{figure}

\subsection*{Excited-state quantum phase transition}

In our final experiment we investigate the excited states of the LMG spectrum.
In the thermodynamic limit the LMG model has an excited-state quantum phase transition (ESQPT) \cite{stransky_excited-state_2014, puebla_non-thermal_2014, cejnar_excited-state_2021} when $h / \gamma_x < 1$, the signature of which is a divergence in its density of states \cite{heiss_large_2005, leyvraz_large-nscaling_2005, ribeiro_thermodynamical_2007, ribeiro_exact_2008}.
We present a semiclassical calculation of the density of states, as well as a discussion of its interpretation, in the Supplemental Material.
In the small-$j$ regime this divergence is replaced by a splitting of the approximate degeneracy between pairs of even- and odd-parity states.
In Figure \ref{fig:excited}a we show the spectrum of the LMG Hamiltonian for $j = 8$.
The red curve indicates the energy difference between the peak of the barrier and the minimum of the semiclassical energy surface as a function of $h / \gamma_x$, which we henceforth call the critical energy $E_c(h)$.
When the eigenenergy is below $E_c(h)$, pairs of even- and odd-parity states are nearly degenerate.
The energies of these states separate when they are roughly equal to $E_c(h)$.
This is shown in Figure \ref{fig:excited}b, where we plot the $j = 8$ spectrum at $h / \gamma_x = 0.18$ on top of the semiclassical energy surface.

We experimentally investigate this behavior by measuring the full energy spectrum of the LMG Hamiltonian for $h / \gamma_x = 0.18$ and $j = 8$.
We begin by noting that, to a very good approximation, the lowest even-odd energy gap is $E_{\ket{1_{\rm LMG}}} - E_{\ket{0_{\rm LMG}}} \approx 0$ for the $h / \gamma_x$ considered here.
This can be inferred from the even-odd gap results in Figure \ref{fig:gap}.
We perform a series of gap measurements similar to those in Figure \ref{fig:gap}, first measuring every even-even gap $E_{\ket{2(n+1)_{\rm LMG}}} - E_{\ket{(2n)_{\rm LMG}}}$, then measuring every odd-odd gap $E_{\ket{(2(n+1) + 1)_{\rm LMG}}} - E_{\ket{(2n + 1)_{\rm LMG}}}$.
From these energy differences, along with the fact that $E_{\ket{1_{\rm LMG}}} - E_{\ket{0_{\rm LMG}}} \approx 0$, we reconstruct the full energy spectrum for $j = 8$ despite the fact that we cannot directly simulate a spin this large.
The protocol, shown in Figure \ref{fig:excited}c and discussed in the Methods section, is analogous to that for the gap measurements in Figure \ref{fig:gap}.

The results of this experiment are shown in Figure \ref{fig:excited}d.
We plot the reconstructed energy splitting for each even-odd pair as a function of the average energy of that pair.
We show the prediction for the splitting as $j \rightarrow \infty$, as well as the semiclassical density of states.
Below the critical energy $E_c(h)$, the even-odd pairs are degenerate for $j \rightarrow \infty$, and nearly degenerate for $j = 8$.
When the energy is above $E_c(h)$ the degeneracy is lifted, and the experimental data for $j = 8$ agree well with the theoretical values for $j = 8$ and $j\to \infty$, where the sharp change in the latter is indicative of an excited-state quantum phase transition.

\section*{Discussion}

In this paper we presented the results of analog quantum simulations of the LMG model in a superconducting transmon qudit using up to $d = 9$ levels, realizing an effective spin of size $j = 4$ or, in cases involving states of definite parity, $j = 8$.
Our implementation of the LMG Hamiltonian is fully tunable, permitting the study of the system's behavior for any (potentially time-dependent) ratio of $h / \gamma_x$, without breaking parity symmetry.
Furthermore, our methods enable the study of ground and excited states, energy gaps, dynamics, order parameters, and other signatures of quantum criticality without prior knowledge of the system's eigenstates.

This experimental progress comes at an opportune time.
In recent years there have been a number of proposals for useful qudit-based quantum simulations, particularly in lattice gauge theory \cite{ciavarella_trailhead_2021, gustafson_prospects_2021, gonzalez-cuadra_hardware_2022, zache_fermion-qudit_2023, illa_qu8its_2024, meth_simulating_2025, calliari_quantum_2025}.
Qudit-based models are also of interest in the areas of quantum chemistry \cite{macdonell_analog_2021} and quantum criticality \cite{pausch_dissipative_2024}.
As demonstrated by our results, the flexible and high-fidelity control afforded to transmon qudits makes them an ideal platform for the experimental realization of these proposals.
The high $E_J / E_C$ ratio of the transmon used in this work is achieved through minor modifications of the device geometry \cite{wang_high-e_je_c_2025}, and it is therefore compatible with state-of-the-art large-scale superconducting processor architectures.
We note that, unlike qudit platforms that make use of natural multi-level quantum systems (e.g., trapped ions \cite{ringbauer_universal_2022, low_control_2025} or nuclear spins \cite{yu_schrodinger_2025, vaartjes_certifying_2025}), a transmon qudit processor can be engineered to optimize the tradeoff between dimensionality and anharmonicity required for a particular simulation \cite{wang_high-e_je_c_2025}.
The connectivity of the processor can also be designed with the simulated Hamiltonian in mind, and recent advances in tailoring interactions between transmon qudits \cite{morvan_qutrit_2021, goss_high-fidelity_2022, nguyen_empowering_2024} suggest that this platform can be scaled to multi-qudit systems.
Our results thus open the door for efficient quantum simulations of complex and practically relevant qudit-based models.

\section{Acknowledgments}
This material is based upon work supported by the Air Force Office of Scientific Research under Award No. FA9550-23-1-0121.
Devices were fabricated and provided by the Superconducting Qubits at Lincoln Laboratory (SQUILL) Foundry at MIT Lincoln Laboratory, with funding from the Laboratory for Physical Sciences (LPS) Qubit Collaboratory.
The traveling-wave parametric amplifier (TWPA) used in this experiment was provided by IARPA and Lincoln Labs.
\section{Methods}

\noindent \textbf{Drive Hamiltonian}.
We write the bare transmon Hamiltonian in the lab frame as
\begin{equation}
    H_0 = \sum_{n=0}^{d-1} \omega_n \ket{n} \bra{n}.
\end{equation}
The lowest transition frequency in our system is \mbox{$(\omega_1 - \omega_0) / 2\pi = \qty{4.870}{\giga\hertz}$}, and the anharmonicity is \mbox{$|\alpha|/2\pi = \qty{104}{\mega\hertz}$}; see the Supplemental Materials for the full list of device parameters.
We drive the transmon with 
\begin{equation}
    H_{\rm drive}(t) = \sum_{k=1}^{d-1} V^{(k)}(t) \sum_{n=1}^{d-1} \gamma_n \bigg ( \ket{n-1} \bra{n} + \ket{n} \bra{n-1} \bigg ),
\end{equation}
consisting of $d - 1$ simultaneous drives $V^{(k)}(t) = \mathcal{E}^{(k)}(t) \cos\left ( \omega^{(k)}(t) t + \phi^{(k)}(t) \right )$, with $\gamma_n$ being the transmon charge matrix elements.
The drive strengths, frequencies, and phases of all drives are potentially time-dependent.
We set each drive frequency $\omega^{(k)}$ to be detuned from the $\ket{k-1} \leftrightarrow \ket{k}$ transition by a small amount: $\omega^{(k)}(t) = (\omega_k - \omega_{k-1}) - (\Delta_k(t) - \Delta_{k-1}(t))$.
Transforming to a rotating frame defined by time-dependent detunings,
\begin{equation}
    R(t) = \exp \left [ i \sum_{n=0}^{d-1} \left ( \omega_n - \int_0^t \diff{t'} \Delta_n(t') \right ) \ket{n} \bra{n} \right ],
\end{equation}
changes the total lab-frame Hamiltonian $H_{\rm lab}(t) = H_0 + H_{\rm drive}(t)$ to 
\begin{equation}
    H_{\rm rot}(t) = R(t) H_{\rm lab}(t) R^\dagger(t) + i \left ( \frac {\diff{}} {\diff{t}} R(t) \right ) R^\dagger(t).
\end{equation}
It is helpful here to define the total accumulated phase relative to the bare transmon frame,
\begin{equation}
    A_n(t) \equiv \int_0^t \diff{t'} \Delta_n(t'),
\end{equation}
as well as the complex-valued drive strength
\begin{equation}
\Omega_n(t) \equiv \frac {1} {2} \gamma_n \mathcal{E}^{(n)}(t) e^{i \phi^{(n)}(t)}.
\end{equation}
We apply the frame transformation and take the rotating wave approximation (RWA) by neglecting terms rotating at the anharmonicity $|\alpha|$ or faster to obtain
\begin{equation}
    \begin{split}
    H_{\rm rot}(t) &= \sum_{n=0}^{d-1} \Delta_n(t) \ket{n} \bra{n} \\
    &+ \sum_{n=1}^{d-1} \bigg ( \Omega_n(t) e^{i \zeta_n(t)} \ket{n-1} \bra{n} + \text{H.c.} \bigg )
    \end{split}
\end{equation}
where
\begin{equation}
    \zeta_n(t) = (\Delta_n(t) - \Delta_{n-1}(t))t - (A_n(t) - A_{n-1}(t)).
\end{equation}
In the case where the detunings are constant, $\zeta_n(t) = 0$.
On the other hand, in the case where the detunings change with respect to time, as they do during the adiabatic ramps, $\zeta_n(t)$ introduces a problematic phase shift on the drives.
We compensate for this by setting the time-dependent phase of each drive to $\phi^{(n)}(t) = -\zeta_n(t)$.
In this way we arrive at the effective Hamiltonian in Eq. \ref{eq:rwa_hamiltonian}, yielding an explicit solution for the drive strengths, frequencies, and phases needed to realize $H_{\rm LMG}$ for any time-dependent $h$ and $\gamma_x$.

\noindent \textbf{Drive generation.} In practice we produce the simultaneous drive tones by modulating a single local oscillator (LO) tone with $d - 1$ fixed-frequency modulation tones via single-sideband mixing.
In what follows we refer to the resulting upconverted frequencies as ``reference frequencies'' because they serve as the phase references for each transition.
When the detunings are constant, $\Delta_n(t) = \Delta_n$, we choose reference frequencies $\beta_n = (\omega_n - \omega_{n-1}) - (\Delta_n - \Delta_{n-1})$.
The frame tracked by these reference frequencies is therefore the one yielding the diagonal elements of Eq. \ref{eq:rwa_hamiltonian}, and the off-diagonal elements are simply given by the strengths of the drives.

The situation is more complicated when the detunings are time-dependent.
In this case, because we cannot smoothly vary the reference frequencies in the hardware, we instead fix them to the values given by the initial detunings: $\beta_n = (\omega_n - \omega_{n-1}) - (\Delta_n(0) - \Delta_{n-1}(0))$.
Recall that the desired waveform is given in the lab frame by
\begin{equation}
    V^{(n)}(t) = \mathcal{E}^{(n)}(t) \cos(\omega^{(n)}(t) t + \phi^{(n)}(t)).
\end{equation}
We emphasize that the time-dependent strengths, frequencies, and phases of these drives are determined by a direct comparison of Eq. \ref{eq:rwa_hamiltonian} to the desired LMG Hamiltonian.
We split each drive tone into a rapidly oscillating component at $\omega^{(n)}(0) = \beta_n$ and a slowly oscillating component:
\begin{equation}
    \begin{split}
        V^{(n)}(t) &= \mathcal{E}^{(n)}(t) \big [ \cos(\beta_n t) \cos(\delta \omega^{(n)}(t) t + \phi^{(n)}(t)) \\
        & \qquad - \sin(\beta_n t) \sin(\delta \omega^{(n)}(t) t + \phi^{(n)}(t)) \big ] \\
        &\equiv \cos(\beta_n t) v_I^{(n)}(t) - \sin(\beta_n t) v_Q^{(n)}(t),
    \end{split}
\end{equation}
where $\delta \omega^{(n)}(t) = \omega^{(n)}(t) - \omega^{(n)}(0) = \omega^{(n)}(t) - \beta_n$.
Therefore, by applying time-dependent envelopes $v_I^{(n)}(t)$ and $v_Q^{(n)}(t)$ to the in-phase and quadrature components of the modulation tones, we realize the desired upconverted signal despite the fact that each reference frequency is, in reality, fixed to a single value for the duration of the pulse.
The final step is to correct the reference frame itself such that the frame tracked by the reference oscillators $\beta_n$ coincides with the desired frame rotation $R(t)$.
This correction amounts to a single instantaneous update of each modulation frequency to its new value at the end of the pulse, as well as a single instantaneous update of the phase of each modulation tone to account for the accumulated phase shift $\phi^{(n)}(t) - \phi^{(n)}(0)$.

\noindent \textbf{Protocols}. In this section we discuss the details of the various experimental protocols used throughout this work.

\begin{enumerate}
    \item \textit{Dynamical phase transition} (Figure \ref{fig:analog_sim}f):
    For the case where the system begins in $\ket{4,4}$ (transmon state $\ket{0}$), we simply turn on the LMG drives as described in the main text for a variable time $t$.
    Following this we measure the qudit state (Supplemental Material) and plot the resulting populations.
    For the case where the system begins in $\ket{4,3}$ (transmon state $\ket{5})$, we apply sequential $\pi$ pulses between pairs of adjacent levels to prepare the initial state prior to turning on the LMG drives.

    \item \textit{Gap measurement} (Figure \ref{fig:gap}):
    Due to the presence of the second-order phase transition at $h / \gamma_x = 1$, we consider the cases where $h / \gamma_x < 1$ and $h / \gamma_x > 1$ separately.
    We note that this protocol can be extended to the case where the critical point is not known (see below).
    The protocol for $h / \gamma_x < 1$ is depicted in Figure \ref{fig:gap}b.
    Beginning in the ground state, we apply sequential pulses between adjacent transmon levels to prepare equal superpositions of the $\ket{j, m}_\pm$ states, defined above as the definite-parity eigenstates of $J_x^2$.
    In particular, for the even-odd gap we prepare $\frac {1} {\sqrt{2}}(\ket{j, j}_+ + \ket{j, j}_-)$, while for the even-even gap we prepare $\frac {1} {\sqrt{2}}(\ket{j, j}_+ + \ket{j, j-1}_+)$.
    Next we turn on the LMG Hamiltonian drives with initial parameters of $\gamma_x = 1$ and $h = 0$, and slowly ramp $h$ to its target value.
    In the limit where this process is adiabatic, the resulting state will be an equal superposition of the instantaneous LMG eigenstates at the end of the ramp.
    We allow the system to evolve for a variable time $\Delta t$ before ramping $h$ back to 0, turning off the LMG drives, and inverting the initial state preparation pulse sequence.
    Finally, we measure the population in the transmon ground state and plot it as a function of time, as shown in Figure \ref{fig:gap}d.
    We note that although the superposition state will accumulate a relative phase during the ramps at a rate given by the instantaneous energy difference between the states, the ramp duration is kept fixed for all time points within a single gap measurement (in this case the ramp duration is $\qty{2}{\micro\second}$).
    The result is an overall phase shift in the oscillation, which does not affect the gap energy estimate.

    The protocol for $h / \gamma_x > 1$, shown in Figure \ref{fig:gap}c, is  analogous to that for $h / \gamma_x < 1$.
    In this case we prepare superpositions of $J_z$ eigenstates, specifically $\frac {1} {\sqrt{2}} (\ket{j,j} + \ket{j,j-1})$ for the even-odd gap and $\frac {1} {\sqrt{2}} (\ket{j,j} + \ket{j,j-2})$ for the even-even gap.
    We turn on the LMG drives, now with $\gamma_x = 0$ and $h$ set to its target value, before slowly ramping $\gamma_x$ to 1.
    We again allow the system to evolve for a time $\Delta t$ before ramping $\gamma_x$ back to 0, inverting the state preparation sequence, and measuring the transmon ground state population.
    In this case, because the energy gaps are larger than when $h / \gamma_x < 1$, the adiabatic approximation is satisfied for faster ramps.
    We therefore choose a ramp duration of $\qty{200}{\nano\second}$.

    If the location of the critical point is not known, one can identify it by applying the protocols given above across the full range of $h / \gamma_x$.
    If the ramp causes the system to cross through the critical point -- i.e., the point where the gap closes -- the system will be excited out of the desired two-state subspace.
    For $h / \gamma_x$ values lying on the same side of the critical point as the initial value, one would observe a single peak in the spectrum.
    For $h / \gamma_x$ values lying on the other side, however, the observation of multiple peaks will indicate the presence of the critical point, and the onset of this behavior should coincide with the measured gap itself reaching a minimum.
    A comparison between the data resulting from the two protocols would therefore give the approximate location of the critical point, and by varying the speed of the ramp and the range of $h / \gamma_x$ values sampled, one could refine the estimate its location.
    In this case, the protocol given here could very well serve as a useful method for locating the critical point.

    \item \textit{Kibble-Zurek} (Figure \ref{fig:ground_state}):
    This protocol begins with the system in the state $\ket{j,j}$, i.e., the ground state of $-J_z$.
    We turn on the LMG drives with $h = 2$ and $\gamma_x$ initially set to 0, then ramp $\gamma_x$ to 1 over the course of $\qty{200}{\nano\second}$.
    Because the gap is large for the duration of this ramp, we faithfully prepare the LMG ground state for $h/\gamma_x = 2$.
    Following this, we ramp $h$ down from 2 to 0 over a variable time $T$.
    In the limit of large $T$ this process will be adiabatic, and so we should have prepared the LMG ground state for $h/\gamma_x = 0$, which due to parity conservation we know to be the $\ket{j, j}_+$ given above.
    We apply sequential pulses to each transition to map this state onto the transmon ground state, measure the transmon ground state population, and plot it as a function of ramp speed $2\pi / (\Omega T)$ in Figure \ref{fig:ground_state}b.

    \item \textit{Statistics of the order parameter} (Figure \ref{fig:order}):

    In Figures \ref{fig:order}f and \ref{fig:order}g we make use of the fact that the LMG ground state has even parity to measure the distribution of $J_x$ eigenstates.
    To do this we note that, defining a projector $P_e$ onto the even subspace, the LMG ground state is unaffected by this projector: $P_e \ket{0_{\rm LMG}} = \ket{0_{\rm LMG}}$.
    The overlap of the LMG ground state with a $J_x$ eigenstate $\ket{j,m}_x$, then, is
    \begin{equation}
        \begin{split}
        |\braket{0_{\rm LMG} | j,m}_x|^2 &= |\braket{0_{\rm LMG} |P_e | j,m}_x|^2 \\
        &= \mathcal{N} {|\braket{0_{\rm LMG} | \psi_m}|^2}
        \end{split}
    \end{equation}
    where $\ket{\psi_m} = \frac{1} {\sqrt{\mathcal{N}}} P_e \ket{j,m}_x$ and $\mathcal{N} = {_x\!\braket{j,m | P_e | j, m}_x}$ is a normalization factor.
    We therefore measure the overlap of $\ket{0_{\rm LMG}}$ with each of the $\ket{\psi_m}$ and renormalize the resulting populations to yield the $J_x$ eigenstate distribution of this larger effective spin.

    \item \textit{Excited-state quantum phase transition} (Figure \ref{fig:excited}):
    This protocol is similar to that for the gap experiments in Figure \ref{fig:gap}.
    The key differences are: (1) we neglect the subspace with the opposite parity from the one we are measuring; (2) we must prepare different initial superpositions prior to the adiabatic ramps; and (3) the adiabatic approximation is more difficult to satisfy because the energy gaps will, in general, become smaller with increasing energy.
    We address this final point by observing that we are free to use different energy scales $\Omega$ for different parts of the protocol.
    In particular, during the adiabatic ramps we increase it to $\Omega / (2 \pi) = \qty{5.730}{\mega\hertz}$ while keeping the ramp duration fixed to $\qty{2}{\micro\second}$.
    This choice has the effect of increasing the energy gaps, permitting a more adiabatic ramp while mitigating the decoherence that would result from simply increasing the ramp duration.
    We note that this strategy comes at the potential cost of more coherent error due to a breakdown of the RWA, and the value of $\Omega$ used here is chosen to balance the various sources of error.
    During the constant-Hamiltonian part of the protocol, meanwhile, the adiabatic condition is not relevant, and we therefore return the energy scale to $\Omega / (2 \pi) =\qty{1.910}{\mega\hertz}$.
\end{enumerate}

\bibliography{references}
\bibliographystyle{naturemag}

\end{document}


\title{Supplemental Material: Analog quantum simulation of the Lipkin-Meshkov-Glick model in a transmon qudit}

\author{Elizabeth Champion}
\author{Annie Schwartz}
\author{Muhammad A. Ijaz}
\author{Xiaohui Xu}
    \affiliation{Department of Physics and Astronomy, University of Rochester, Rochester, New York 14627, USA}
    \affiliation{University of Rochester Center for Coherence and Quantum Science, Rochester, New York 14627, USA}
\author{Steve Campbell}
    \affiliation{School of Physics, University College Dublin, Belfield, Dublin 4, Ireland}
    \affiliation{Centre for Quantum Engineering, Science, and Technology, University College Dublin, Dublin 4, Ireland}
\author{Gabriel T. Landi}
\author{Machiel Blok}
    \email{machielblok@rochester.edu}
    \affiliation{Department of Physics and Astronomy, University of Rochester, Rochester, New York 14627, USA}
    \affiliation{University of Rochester Center for Coherence and Quantum Science, Rochester, New York 14627, USA}

\date{\today}

\maketitle

\beginsupplement


\onecolumngrid

\tableofcontents

\section{High-$E_J/E_C$ transmon device}

\begin{figure*}
    \includegraphics{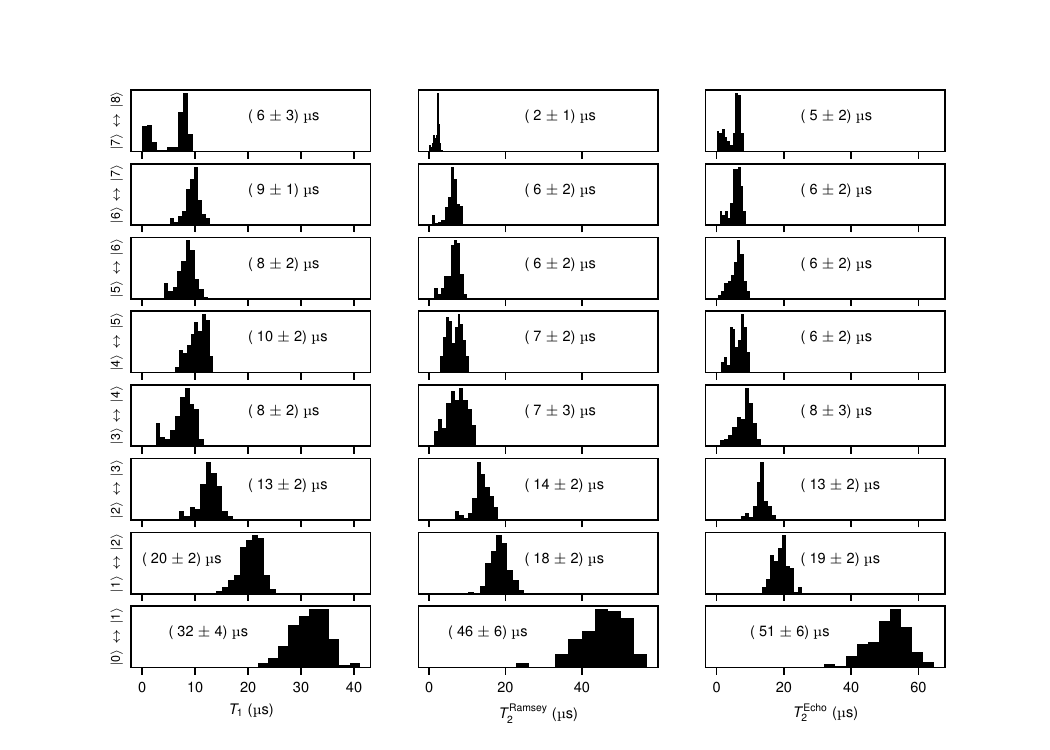}
    \caption{Histograms of measured coherence times for 120 repetitions over the course of roughly 20 hours.
    The numerical values reported in each panel are means and standard deviations.}
    \label{fig:coherences}
\end{figure*}

The experiments presented in this work employed a superconducting qudit device with sufficiently many accessible levels to emulate the large spin.
For this purpose we use a transmon qudit with a large ratio of $E_J / E_C \approx 300$, where $E_J$ is the Josephson energy and $E_C$ is the capacitive energy \cite{koch_charge-insensitive_2007, blais_circuit_2021}.
We provide the device parameters in Table \ref{tab:device}.
We perform single-shot readout of the qudit using a multi-tone readout protocol in which a dispersively coupled resonator is probed at three frequencies simultaneously \cite{wang_high-e_je_c_2025, champion_efficient_2025}.
In Figure \ref{fig:coherences} we report measured values for $T_1$, $T_2^{\rm Ramsey}$, and $T_2^{\rm Echo}$ for every transition used in this work.
Each coherence time is measured 120 times over the course of roughly 20 hours, and we show histograms of the measured times; the numerical values reported in the Figure are the means and standard deviations across this data set.

We refer the reader to Ref. \cite{wang_high-e_je_c_2025}, in which the transmon used here is labeled $Q_5$, for an in-depth discussion of the device, its noise properties, and the multi-tone readout protocol.
We note that the experiments in this work were performed during a different cooldown from those in Ref. \cite{wang_high-e_je_c_2025}, which accounts for the slight discrepancies between the device parameters and coherence times reported here and those reported in earlier work.

\begin{table*}
\begin{center}
\setlength{\tabcolsep}{4pt}
\begin{tabular}{ c c c c c c c c | c c c c }
    \toprule
    $f_{01}$ & $f_{12}$ & $f_{23}$ & $f_{34}$ & $f_{45}$ & $f_{56}$ & $f_{67}$ & $f_{78}$ & $E_J$ & $E_C$ & $f_r$ & $g$ \\
    \midrule
    4.870 & 4.766 & 4.658 & 4.544 & 4.424 & 4.298 & 4.163 & 4.020 & 30.497 & 0.102 & 6.471 & 0.029 \\
    \bottomrule
\end{tabular}
\caption{Measured transition frequencies and calculated $E_J$, $E_C$, bare resonator frequency $f_r$, and coupling strength $g$. All values are expressed in GHz.}
\label{tab:device}
\end{center}
\end{table*}

\section{LMG eigenstates}

The LMG Hamiltonian we simulate in this work is given by
\begin{equation}
    H_{\rm LMG} / \Omega = -h J_z - \frac {\gamma_x} {2j} J_x^2.
\end{equation}
We work in the basis of $J_z$ eigenstates under the labeling scheme presented in the main text.
The $J_z$ eigenstates for a spin-$j$ system are denoted by $\ket{j,m}$, where
\begin{equation}
    J_z \ket{j,m} = m \ket{j,m}.
\end{equation}
The eigenstates of $J_x$, meanwhile, are denoted by $\ket{j,m}_x$, where
\begin{equation}
    J_x \ket{j,m}_x = m \ket{j,m}_x.
\end{equation}
These bases are related by a $\pi/2$ rotation about $J_y$: rewriting $J_x$ as $e^{-i \frac {\pi} {2} J_y} J_z e^{i \frac {\pi} {2} J_y}$ we have
\begin{equation}
    \left ( e^{-i \frac {\pi} {2} J_y} J_z e^{i \frac {\pi} {2} J_y} \right ) \ket{j,m}_x = m \ket{j,m}_x,
\end{equation}
implying that
\begin{equation}
    J_z \left ( e^{i \frac {\pi} {2} J_y} \ket{j,m}_x \right ) = m \left ( e^{i \frac {\pi} {2} J_y} \ket{j,m}_x \right ),
\end{equation}
so we see that
\begin{equation}
    \ket{j,m}_x = e^{-i \frac {\pi} {2} J_y} \ket{j,m}.
\end{equation}

We define the parity operator used throughout the main text as 
\begin{equation}
    \Pi = e^{i \pi (J_z - j)},
\end{equation}
which has the properties $\Pi^2 = \mathbbm{1}$ and $\Pi^\dagger = \Pi$.
Because $J_x^2$ commutes with $\Pi$, they share a simultaneous eigenbasis, and we therefore seek the definite-parity eigenstates of $J_x^2$.
Noting that $\Pi J_x \Pi^\dagger = -J_x$, we have 
\begin{equation}
    \begin{split}
        \Pi J_x \ket{j,m}_x &= m \Pi \ket{j,m}_x \\
        \Pi J_x ( \Pi^\dagger \Pi ) \ket{j,m}_x &= m \Pi \ket{j,m}_x \\
        -J_x \Pi \ket{j,m}_x &= m \Pi \ket{j,m}_x,
    \end{split}
\end{equation}
so we identify
\begin{equation}
    \Pi \ket{j,m}_x = \ket{j,-m}_x.
\end{equation}
The definite-parity eigenstates of $J_x^2$ thus have the form $\alpha \ket{j,m}_x + \beta \ket{j,-m}_x$.
Substituting we find
\begin{equation}
    \Pi (\alpha \ket{j,m}_x + \beta \ket{j,-m}_x) = \alpha \ket{j,-m}_x + \beta \ket{j,m}_x,
\end{equation}
requiring $\alpha = \pm \beta$ (except when $m = 0$).
We therefore define
\begin{equation}
    \ket{j,m}_\pm = \frac {1} {\sqrt{2}}(\ket{j,m}_x \pm \ket{j,-m}_x)
\end{equation}
when $m \neq 0$, and
\begin{equation}
    \ket{j,0}_+ = \ket{j,0}_x.
\end{equation}
It is clear, then, that
\begin{equation}
    J_x^2 \ket{j,m}_\pm = m^2 \ket{j,m}_\pm
\end{equation}
and
\begin{equation}
    \Pi \ket{j,m}_\pm  = \pm \ket{j,m}_\pm.
\end{equation}

\section{Semiclassical model}

In this section we summarize the semiclassical approach to modeling properties of the LMG Hamiltonian.
We refer the reader to Refs. \cite{leyvraz_large-nscaling_2005, castanos_classical_2006, ribeiro_exact_2008, ribeiro_thermodynamical_2007} for more thorough discussions of this approach and its applications.

\subsection{Classical energy surface and ground state solutions}

The semiclassical model discussed in the main text, and used below for the density of states calculation, is obtained by noticing that, in the thermodynamic limit, the angular momentum operators $(J_x, J_y, J_z)$ can be replaced by a vector lying on a sphere of radius $j$:
\begin{equation}
    (J_x, J_y, J_z) \leftrightarrow j(\sin\theta \cos\phi, \sin\theta \sin\phi, \cos\theta).
\end{equation}
Substituting these into the LMG Hamiltonian we find
\begin{equation}
    \mathcal{H}(\theta, \phi) = \frac {E(\theta, \phi)} {j} = -h \cos\theta - \frac {\gamma_x} {2} \sin^2\theta \cos^2\phi.
\end{equation}

The same energy surface is also found by evaluating the expectation value of the Hamiltonian for a spin coherent state, $\ket{\theta, \phi} = e^{-i \phi J_z} e^{-i \theta J_y} \ket{j,j}$, in the limit $j \rightarrow \infty$.
First, note that the taking the expectation value of each angular momentum operator gives
\begin{equation}
    (\langle J_x \rangle, \langle J_y \rangle, \langle J_z \rangle) = j(\sin\theta \cos\phi, \sin\theta \sin\phi, \cos\theta).
\end{equation}
For large $j$ we have $\langle \theta, \phi | J_\alpha^2 | \theta, \phi \rangle \approx \langle \theta, \phi | J_\alpha | \theta, \phi \rangle^2$ where $\alpha = x, y, z$, and in the limit $j \rightarrow \infty$ the relation becomes exact. 
We therefore recover the energy surface given above,
\begin{equation}
    \mathcal{H}(\theta, \phi) = \lim_{j \rightarrow \infty} \frac {1} {j} \langle \theta, \phi | H | \theta, \phi \rangle.
\end{equation}

The ground states of the semiclassical model are found by differentiating $\mathcal{H}$ with respect to $\theta$ and $\phi$ and setting the result equal to zero.
Carrying out this calculation, one finds that when $h > \gamma_x$, there is a single minimum at $\theta_0 = 0$, with $E_{\rm min} = \mathcal{H}(0, \phi) = -h$.
When $h < \gamma_x$, however, there are two distinct minima with $\phi_0 = 0, \pi$ and $\cos\theta_0 = \frac {h} {\gamma_x}$, with degenerate energies
\begin{equation}
    E_{\rm min} = \mathcal{H}(\theta_0, 0) = \mathcal{H}(\theta_0, \pi) = - \frac {h^2 + \gamma_x^2} {2 \gamma_x}.
\end{equation}
We note that, for energy spectra plotted in the main text, we shift all energies such that the ground state energy is zero for all $h / \gamma_x$, which aids comparisons with the experimental results.

\subsection{Density of states}

In this section we present a calculation of the LMG density of states in the semiclassical approximation.
The essential aspects of the calculation can be found in, e.g., Ref. \cite{nader_avoided_2021}.
Here we elucidate the details, as they provide valuable insights into the origin of the excited-state quantum phase transition and the connections between the quantum Hamiltonian and the semiclassical model.

For convenience, here we set $\gamma_x = 1$ and define $z = -\cos\theta$, so that the semiclassical Hamiltonian is given by
\begin{equation}
    \mathcal{H}(z, \phi) = h z - \frac {1 - z^2} {2} \cos^2\phi.
\end{equation}
The density of states at an energy $E$ is approximated semiclassically by the Gutzwiller trace formula:
\begin{equation} \label{eq:traceformula}
    \rho(E) = \frac {1} {4 \pi} \int \dd{z} \dd{\phi} \delta[\mathcal{H}(z,\phi) - E].
\end{equation}
One can loosely imagine that this integral counts the number of points in the classical phase space with a given energy $E$.
The normalization factor of $4 \pi$ is obtained by integrating Eq. \ref{eq:traceformula} over all values of $E$.
We will approach this integral by first finding the $z$ and $\phi$ which yield $\mathcal{H}(z,\phi) = E$ for a given $E$.
It is helpful to note that, in the region $h \leq 1$, $E$ is bounded from below by $E \geq -\frac {h^2 + 1} {2}$, while when $h \geq 1$, $E \geq -h$.
In both cases, $E$ is bounded from above by $E \leq h$.

For a given $z$ and $E$, $\phi$ is determined by
\begin{equation}
    \phi = \pm \cos^{-1} \left ( \pm \sqrt{2 \frac {h z - E} {1 - z^2}} \right ),
\end{equation}
which only has solutions when
\begin{equation}
    0 \leq 2 \frac {h z - E} {1 - z^2} \leq 1.
\end{equation}
From this it is straightforward to show that, when $h \leq 1$,
\begin{equation}
    -h - \sqrt{h^2 + 2 E + } \leq z \leq -h + \sqrt{h^2 + 2 E + 1},
\end{equation}
and when $h \geq 1$,
\begin{equation}
    \frac {E} {h} \leq z \leq -h + \sqrt{h^2 + 2 E + 1}.
\end{equation}

\begin{figure*}
    \includegraphics{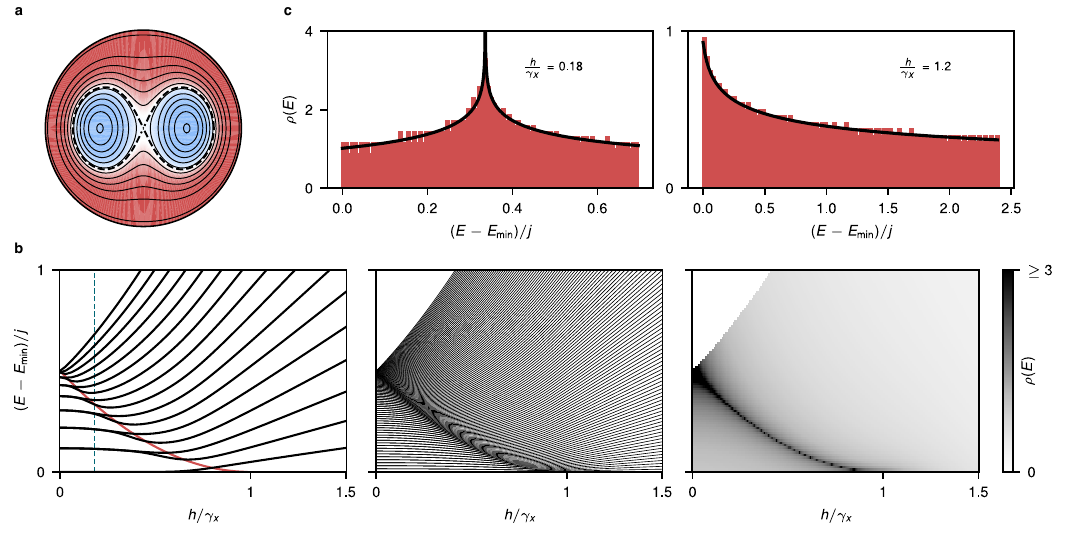}
    \caption{Density of states of the LMG model.
    \textbf{a} The semiclassical energy surface $\mathcal{H}(\theta, \phi)$, represented in color, with constant-energy curves plotted on top.
    We use a polar projection of the sphere, where the radial coordinate is $\theta$ and the angular coordinate is $\phi$; in this mapping the north pole of the sphere corresponds to the center of the disk while the south pole corresponds to its boundary.
    \textbf{b} Energy spectra for $j = 8$ (left) and $j = 64$ (middle) and the semiclassical density of states (right).
    The red curve in the left panel is the critical energy of the ESQPT, and the vertical dashed line indicates the value of $h / \gamma_x$ used in both the ESQPT experiments in the main text and in \textbf{a}.
    \textbf{c} Semiclassical density of states plotted on top of histograms of eigenenergies for $j = 500$.}
    \label{fig:dos}
\end{figure*}

On the other hand, for a given $\phi$ and $E$, $z$ takes one or both roots
\begin{equation} \label{eq:zroots}
    z_\pm = \frac {-h \pm \sqrt{h^2 + \cos^2\phi \left ( 2 E + \cos^2\phi \right )}} {\cos^2\phi}.
\end{equation}
First consider the case where $h \leq 1$.
When $E < -h$, both values in Eq. \ref{eq:zroots} yield a valid $z \in [-1, 1]$.
When $E > -h$, meanwhile, only the $z_+$ root is valid.
This change is closely related to the excited state quantum phase transition: below the transition there are two distinct classical trajectories with the same $E$, while above, there is only one.
This is depicted in Figure \ref{fig:dos}a, where we show curves of constant energy on top of the classical energy surface for $h = 0.18$.
For energies below the transition, the constant-energy curves come in pairs.
For $h \geq 1$, meanwhile, there is always only one valid solution, namely, the $z_+$ root.

To evaluate the integral in Eq. \ref{eq:traceformula}, we must also find the set of $\phi$ values for which there is at least one $z$ such that $\mathcal{H}(z, \phi) = E$ for a given $E$.
When $h \geq 1$, this is simply $\phi \in [0, 2\pi]$.
We also have $\phi \in [0, 2\pi]$ for $h \leq 1$ when $E \geq -h$.
The case where $h \leq 1$ and $E \leq -h$ is somewhat more complicated.
Here $\phi$ takes values on two disjoint intervals:
\begin{equation}
    \phi \in \left [ -\phi_0, \phi_0 \right ]
\end{equation}
and \begin{equation}
    \phi \in \left [ \pi - \phi_0, \pi + \phi_0 \right ]
\end{equation}
where in both cases
\begin{equation}
    \phi_0 = \cos^{-1}\left ( \sqrt{-h \zeta_-}  \right )
\end{equation}
and
\begin{equation}
    \zeta_\pm = \frac {E} {h} \pm \sqrt{\left ( \frac {E} {h} \right )^2 - 1}
\end{equation}
(we will make use of $\zeta_+$ shortly).
In every case we denote the set of permissible $\phi$ values by $\bm{\Phi}_E$.

We now proceed with the integral in Eq. \ref{eq:traceformula}.
\begin{equation}
    \rho(E) = \frac {1} {4 \pi} \int_{\phi \in \bm{\Phi}_E} \dd{\phi} \int_{-1}^1 \dd{z} \delta[g_{E,\phi}(z)]
\end{equation}
where
\begin{equation}
    g_{E,\phi}(z) \equiv \mathcal{H}(z, \phi) - E.
\end{equation}
We apply the following property of the Dirac delta:
\begin{equation}
    \int_{-\infty}^\infty \dd{x} f(x) \delta[g(x)] = \sum_i \frac {f(x_i)} {|g'(x_i)|}
\end{equation}
where the $x_i$ are the roots of $g(x)$.
In our case the roots of $g(z)$ are the $z_\pm$ we found above, and we have
\begin{equation}
    g'(z_\pm) = \pm \sqrt{h^2 + \cos^2\phi \left ( 2 E + \cos^2\phi \right )}.
\end{equation}
We evaluate the integral in the following three regions: Region I, where $h \leq 1$ and $E \leq -h$; Region II, where $h \leq 1$ and $E \geq -h$; and Region III, where $h \geq 1$.
Starting with Region I:
\begin{equation}
    \rho_{\rm I}(E) = \frac {1} {4\pi} \int_{-\phi_0}^{\phi_0} \frac {2\dd{\phi}} {\sqrt{h^2 + \cos^2\phi \left ( 2 E + \cos^2\phi \right )}} + \frac {1} {4\pi} \int_{\pi-\phi_0}^{\pi+\phi_0} \frac {2\dd{\phi}} {\sqrt{h^2 + \cos^2\phi \left ( 2 E + \cos^2\phi \right )}},
\end{equation}
which can be rewritten as
\begin{equation}
    \rho_{\rm I}(E) = \frac {2} {\pi} \int_0^{\phi_0} \frac {\dd{\phi}} {\sqrt{h^2 + \cos^2\phi \left ( 2 E + \cos^2\phi \right )}}.
\end{equation}
We will evaluate this integral numerically, but we must first deal with the fact that the integrand has a singularity at $\phi = \phi_0$.
Our approach will be to subtract a function from the integrand that has the same asymptotic behavior as $\phi \rightarrow \phi_0$, and that has an analytic solution, such that we can add its contribution back after the numerical integration.
We first factor the denominator:
\begin{equation}
    \rho_{\rm I}(E) = \frac {2} {\pi} \int_0^{\phi_0} \frac {\dd{\phi}} {\sqrt{(\cos^2\phi + h \zeta_+)(\cos^2\phi + h \zeta_-)}}.
\end{equation}
The origin of the singularity with $\phi = \phi_0 = \cos^{-1}(\sqrt{-h \zeta_-})$ is now clear.
The function
\begin{equation}
    \frac {1} {\sqrt{(\cos^2\phi_0 + h \zeta_+)}} \frac {1} {\sqrt{(\cos^2\phi + h \zeta_-)}} \equiv \frac {s(\zeta_+, \zeta_-)} {\sqrt{(\cos^2\phi + h \zeta_-)}},
\end{equation}
where $s(\zeta_+, \zeta_-)$ is constant with respect to $\phi$ has the same asymptotic behavior as $\phi \rightarrow \phi_0$ as our integrand.
Then
\begin{equation}
    \rho_{\rm I}(E) = \frac {2} {\pi} \int_0^{\phi_0} \dd{\phi} \left \{ \frac {1} {\sqrt{(\cos^2\phi + h \zeta_+)(\cos^2\phi + h \zeta_-)}} - \frac {s(\zeta_+, \zeta_-)} {\sqrt{(\cos^2\phi + h \zeta_-)}}\right \} + \frac {2} {\pi} \int_0^{\phi_0} \dd{\phi} \frac {s(\zeta_+, \zeta_-)} {\sqrt{(\cos^2\phi + h \zeta_-)}}.
\end{equation}
The first integral is easily evaluated numerically, as we have in effect ``subtracted away'' its singularity.
The second integral, meanwhile, can be evaluated as
\begin{equation}
    r(\zeta_+, \zeta_-) = \frac {2} {\pi} \int_0^{\phi_0} \dd{\phi} \frac {s(\zeta_+, \zeta_-)} {\sqrt{(\cos^2\phi + h \zeta_-)}} = \frac {2} {\pi} s(\zeta_+, \zeta_-) \sqrt{ \frac {1} {1 + h \zeta_-}} F \left ( \phi_0, \frac {1} {1 + h \zeta_-} \right )
\end{equation}
where $F(\phi, m)$ is the incomplete elliptic integral of the first kind.
This contains a slight problem in that $m$ here will generally be greater than 1, causing problems for the numerical evaluation of this expression.
We can use the following identity to rectify this problem:
\begin{equation}
    F(\phi, m) = \frac {1} {\sqrt{m}} F\left ( \sin^{-1}(\sqrt{m} \sin\phi), \frac {1} {m} \right ).
\end{equation}
Substituting $\phi = \phi_0$ and $m = \frac {1} {1 + h \zeta_-}$ we obtain
\begin{equation}
    F \left ( \phi_0, \frac {1} {1 + h \zeta_-} \right ) = F \left ( \frac {\pi} {2}, 1 + h \zeta_- \right ),
\end{equation}
which is equal to the complete elliptic integral of the first kind, $K(1 + h \zeta_-)$.
Therefore,
\begin{equation}
    r(\zeta_+, \zeta_-) = \frac {4} {\pi} s(\zeta_+, \zeta_-) K(1 + h \zeta_-).
\end{equation}
The density of states in Region I, then, is ultimately given by
\begin{equation}
    \rho_{\rm I}(E) = \frac {2} {\pi} \int_0^{\phi_0} \dd{\phi} \left \{ \frac {1} {\sqrt{(\cos^2\phi + h \zeta_+)(\cos^2\phi + h \zeta_-)}} - \frac {s(\zeta_+, \zeta_-)} {\sqrt{(\cos^2\phi + h \zeta_-)}}\right \} + r(\zeta_+, \zeta_-),
\end{equation}
where the integral is easily evaluated by standard numerical methods.

In Regions II and III the integrand is well-behaved, and noting that in both cases there is only one root of $g(z)$, the density of states is given by a numerical integral of
\begin{equation}
    \rho_{\rm II}(E) = \rho_{\rm III}(E) = \frac {1} {4\pi} \int_0^{2\pi} \frac {\dd{\phi}} {\sqrt{h^2 + \cos^2\phi \left ( 2 E + \cos^2\phi \right )}}.
\end{equation}

In Figure \ref{fig:dos}b we show the energy spectrum for $j = 8$ (left) and $j = 64$ (middle), as well as the density of states as calculated here (right).
The increased density of states near the critical energy is visible in the energy spectra, particularly for $j = 64$.
This density diverges in the thermodynamic limit, and this divergence coincides with the splitting of even- and odd-parity pairs near the critical energy.
Finally, in Figure \ref{fig:dos}c we quantitatively validate our calculation by plotting the semiclassical density of states for two values of $h$ on top of a histogram of the respective eigenenergies for $j = 500$, obtained by numerical diagonalization.

\bibliography{references}
\bibliographystyle{naturemag}